# A Random Access Reconfigurable Metamaterial and a Tunable Flat Lens

W. M. Zhu[1], Q. H. Song[2], A. Q. Liu[1,†], D. P. Tsai[3], H. Cai[4], Z. X. Shen[1], R. F. Huang[5], S. K. Ting[5], Q. X. Liang[2], H. Z. Liu[2], B. H. Lu[2], and N. I. Zheludev[6,7,†]

[1]School of Electrical and Electronic Engineering
Nanyang Technological University, Singapore 639798
[2]School of Mechanical Eengineering, Xi'an Jiangtong University, Xi'an 710049, China
[3]Department of Physics, National Taiwan University, Taiwan 10617
[4]Institute of Microelectronics, A*START, Singapore 117686
[5]Temasek Laboratories, 5A Engineering Drive 1, Singapore 117411
[6]Optoelectronics Research Centre, Southampton SO17 1BJ, United Kingdom
[7]Centre for Disruptive Photonic Technologies, School of Physical and Mathematical Sciences
Nanyang Technological University, Singapore 637371
([†] Email: eaqliu@ntu.edu.sg )

**The ability to control resonant properties of individual metamolecule in a metamaterial structure will offer an ultimate freedom for dynamically shaping wavefronts of electromagnetic radiation for applications such as variable aberration corrected planar lenses, dynamic holograms and spatial intensity and phase modulators. Here we report the first demonstration of a metamaterial where resonant properties of every individual metamolecule can be continuously controlled at will thus offering an ultimate freedom in achieving a dynamic control of electromagnetic wavefront of microwave radiation. We call this a Random Access Reconfigurable Metamaterial (RARM). To achieve a RARM we created an array of cavities that were filled with liquid metal in a precise fashion using microfluidic technology. The developed RARM was used provide the first demonstration of a tunable flat lens.**

The concept of metamaterials, electromagnetic media structured on sub-wavelength scale, has created a platform of new opportunities for manipulating light across the entire electromagnetic spectrum. Metamaterial with manageable dispersion allows access to unusual permittivities and permeabilities leading to negative index [1-2] zero epsilon [3] giant chirality [4] or exotic and useful hyperbolic dispersion anisotropy [5]. Metamaterials can form "invisible" metallic structures and exhibit extraordinary resonant transparency [6-9]. Controlling boundary conditions with metamaterial offers perfect absorbing media [10, 11] and "magnetic" mirrors [12-13]. Metamaterials are now widely exploited to enhance nonlinear, switching and light emission [14, 15] performance of conventional active materials. Metamaterials allow waveform manipulation [16] and offer exciting opportunities for cloaking [17, 18], waveguiding [19] and localization of light. Moreover, metamaterials are now used as a platform for exploration and modelling of new physical effects [20, 21] and developing practical sensor solutions [22, 23]. These thrilling technological prospects have stimulated a wide search for developing metamaterials with tunable and switchable properties using MEMS, phase change media, liquid crystal, magnetic media and superconductors [24, 25]. Efficient modulation of reflection and transmission of metamaterial array in the terahertz and sub-terahertz parts of the spectrum can also be achieved by injecting current into the supporting semiconductor substrate [26] or into a wire loop continuously connecting all metamolecules [27]. Controlling physical shape or mutual position of metamolecules in metamaterial arrays also allows for a very efficient tuning of their characteristics, which can now be achieved with megahertz bandwidth in the optical part of the spectrum [28].

Researchers are now exploring planar metamaterials with spatially variable characteristics as diffraction grating [29-33] and for focusing [34, 35] and super-focusing applications [36]. The ability to control resonant properties of every individual metamolecule in a structure that we call a Random Access Reconfigurable Metamaterial (RARM) will offers an ultimate freedom for dynamically shaping wavefronts for applications such as variable aberration corrected planar lenses, dynamic holograms and spatial intensity and phase modulators with sub-wavelength pixilation. RARMs are passive structures that are designed to control wavefront of external source of radiation. They shall be distinguished from Phased Antenna Arrays that are active structures shaping radiated pattern by controlling phase and intensity characteristics of every emitter through a complex feed arrangement.

Techniques based on integrating lump active elements in the metamolecules (transistors or diodes), current injection in semiconductor substrate, or suppression of superconductivity are potentially suitable for controlling individual metamolecules in RARMs, but depend for modulation on undesirable Joule losses. Moreover, to implement a two-dimensional control of the array they will require a network of wires, individual electrical connections to all metamolecules, which will inevitably interfere and could spoil electromagnetic resonant properties of the metamaterial. Here we report for the first time a proof-of-principle demonstration of a planar RARM where resonant properties of every individual metamolecule can be continuously controlled at will. Using the developed RARM we also provide the first demonstration of a metamaterial lens that is tuneable in focal distance and repositionable spatially at will and shows a multiple foci function. To achieve RARM we have created an array of cavities that can be filled with liquid metal [37-41] in a controllable fashion using microfluidic technology and pneumatic control. A microfluidic network addressing every individual element of the array provides the mechanism to dynamically change the filling factor of resonators and thus their resonant electromagnetic properties at will, continuously and with random access. Such individually addressable cavities of sub-wavelength size form a metamaterial array of sub-wavelength resonators where modulation of transmitted wavefront is achieved practically without Joule losses [42-43] via balancing reflection and transmission of the array. Our metamaterial does not have massive moving parts characteristic to comb-driven MEMS metamaterial [44-46] while by using pneumatic control the proposed solution minimizes the network of conductive elements that could disturb electromagnetic properties of the array and can be used with metamolecules of broad variety of shapes, from simple antennas to complex connected and disconnected three dimensional structures. Our proof-of-principle demonstration is based on microfluidic elements allowing RARMs operating in microwave regime. However, with recent progress in nano-fluidics (for instance, the flow of liquids through carbon nanotubes is being investigated [47]) one can envisage a realization of nanocapillary RARM that could translate the concept into higher frequencies, including the optical range.

The metamaterial reported here is a two-dimensional square array of metallic split rings (Fig. 1a). The split ring element has a radius of 2 mm in a $9 \times 9$ lattice with periodic spacing of 7 mm, which has a total footprint of 63 mm. It is designed to operate in the GHz range from 12 to 18 GHz ($K_u$ band). The split ring metamolecules are formed by filling liquid mercury into the ring-shaped micro cavities. If completely filled with mercury the ring metamolecule exhibits a dipole absorption resonance at the wavelength linked to the half-length of the ring. The resonance wavelength can be progressively reduced if a section of the ring is removed by introducing a gap into the ring. This can simply be achieved by substituting the liquid metal in the cavity with a gas bubble. To create and control the gaps of the rings individually, they are connected by micro channels to pneumatic valves regulating air pressure in the channels. By increasing pressure in the channel air pushes mercury away from the cavity substituting it with an air bubble. The process can be made continuous and fully reversible by changing the balance of air and mercury pressures. Here mercury is chosen for its low melting point (-38.8 ℃) and high electrical conductivity ($1.04 \times 10^6$ S/m), which is only one order of magnitude lower than in copper ($5.96 \times 10^7$ S/m). The metamaterial is supported on a Pyrex glass substrate of 1-mm thick while the architecture of micro-channels and cavities is imbedded into a polydimethylsiloxane layered structure of 1-cm thick. Ring cavities are located in the layer bonded with the glass substrate that also hosts mercury filling channels. Mercury and air channels predominantly extend along the direction perpendicular to the plane of symmetry of the cavity with bubble. This minimizes their influence on the resonance properties of the array for the electromagnetic radiation polarized along the symmetry axis. Narrow air channels directed across the rings connect each cavity with the vertical air loading channels that are linked to the pattern of pneumatic valves and air inlet in the layers above, as shown in Fig. 2a. The control system is fabricated within a polydimethylsiloxane layered structure of 3-cm thick. In the heart of the air control system is a purposely designed ternary valve multiplexer addressing $3^4 = 81$ metamolecules in the array individually (see details in the supplementary materials Fig. S1).

We illustrate the opportunities provided by the flexibility of random access metamaterial by functionalizing it into a tunable flat lens (Fig. 1b). Such a lens can be created by altering the shape of a single metamolecule in the array. As excitations induced by the incident wave in the metamolecules of the array are dipole-coupled, resonant properties of the metamolecules surrounding the altered one are also affected. Indeed, the results of 3D full Maxwell numerical modelling of the uniform array with one metamolecule tuned off by changing the metal filling factor show the lens formation (Fig. 3b & Fig. 3c). Here we assume that the incident electric field is parallel to the symmetry axis of the ring resonator within the *xy* plane. Figure 3c shows the phase profile when the incident wave of 15 GHz passed through the homogeneous array when all the gaps of the metamolecules are identical (gap opening is 5°). Here the wavefront is slightly bended at the edges only due to the finite size of the array. However, if the incident wave passes through the array that contains one altered metamolecule with wider gap opening (12.5°), the wavefront converges, as shown in Fig. 3b. Numerical analysis on phase delays induced to the transmitting wave by the metamolecules array with different gaps shows that at frequencies lower than the resonance frequency of an undisturbed array (18.5 GHz) the phase change inflicted by the disturbed ring will increase monotonously with the gap in the ring, Fig. 3d. The inserts show that no abrupt mode change occurs during the alteration of the gap size. Moreover, the amplitude of the transmitted wave could be well maintained within a small range of gap tuning at the low dispersive region. For the purpose of demonstrating a convex lensing function the incident wave frequency was chosen at 15 GHz.

Experimental results of electromagnetic wave focusing with RARM are presented on Fig. 4. Here the metamaterial is illuminated by a plane wave that propagates along *z*-direction normal to the plane of the metamolecules array and is polarized along the symmetry axis *x* of the metamolecules. All measurements are performed in microwave anechoic chamber using a vector network analyzer with horn antenna as the wave source and a monopole antenna mounted on an *xy* scanner as a probe. To map the transmitted field, the receiving probe is scanned at different *z* distances from the sample with steps of 1.5 mm. An aluminium shutter with a 9 cm ×9 cm window is used to filter the incident electromagnetic wave far from the edge of the RARM. The electric field amplitude distribution at the hotspots is shown as 2D contour map in *xz* plane and *xy* plane in the left and middle column, respectively. The *xz* and *xy* cross-sections illustrate that the beam can be confined into an intensity hotspot with FWHM = 1.81 λ when the ring gap of the central, disturbed metamolecule is 17°, while for the rest of the array it is kept at 5° (Fig. 4). It is possible to further enhance the RARM lens's resolution by improving its numerical aperture. This could be achieved by using larger arrays and providing gradient gap change, affecting more meta-molecules. Figs. 4a-d show that the focal length *F* can be tuned from 2.4 λ to 5.3 λ by adjusting the gap of the central metamolecule from 17° to 11°. Here, the focal length is defined as the distance between the central of the hotspot and the surface of the PDMS layer. In general case, the lens is slightly anisotropic with respect to the polarization state of light with main anisotropic directions to be parallel and perpendicular to the symmetry axis of the ring resonator, which leads to a cylindrical aberration of the lens and results in elliptical hotspot (Fig. 4e-h). Fig. 4i-l show the corresponding numerical analysis of the intensity distribution at *xy* plane across the centre of the hotspot (see details in the supplementary materials Fig. S2). The cylindrical aberration is due to the dipole coupling between the central and the neighbouring meta-molecules, which is minimized when resonance of the central meta-molecule is weak. Fig. 4c shows the distorted hotspot when the gap openings of the central metamolecules are not optimized.

In conclusion, a random access metamaterial formed by casting liquid metal microwave resonators through microfluidic channels provides a flexible platform for wavefront manipulation, which has been illustrated by showing a tunable lens. The RARM can be used as densely integrated tunable lens array, which has potential applications in high resolution display, sensor and imaging systems. We expect that in the future random access reconfigurable metamaterials will be developed for the entire electromagnetic spectral range up to optical frequencies. This will make possible various applications such as 3D holographic displays for mobile phones, high-performance devices for space division multiplexing in the next generation telecommunication networks, and adaptive wavefront correction devices, to name just a few.

**Acknowledgements**
The work is main supported by the Environmental and Water Industry Development Council of Singapore (EWI), RPC programme (Grant No. 1102-IRIS-05-01 and 1102-IRIS-05-02); Ministry of Education (MOE) (Grant No. MOE2011-T3-1-005) and EPSRC (UK) Programme on Nanostructured Photonic Metamaterials; National Science Council of Taiwan (Grant No. NSC 101-2811-P-002-004 and NSC 102-2745-M-002-005-ASP) and National Natural Science Foundation of China (Grant No. 91223203).

**Author contributions**
W. M, Z., N. I. Z, D. P. T. and A. Q. L. jointly conceived the idea. Q. H. S. and H. C. assisted in the fabrication and figure preparation. R. F. H., Z. X. S. and S. K. T. assisted in the experiment. Q. X. L., H. Z. L. and B. H. L. assisted in the analysis and discussion of the results. W. M. Z., N. I. Z., D. P. T. and A. Q. L. prepared the manuscript. A. Q. L. supervised and coordinated in the all work.

**Additional information**
The authors declare no competing financial interests.

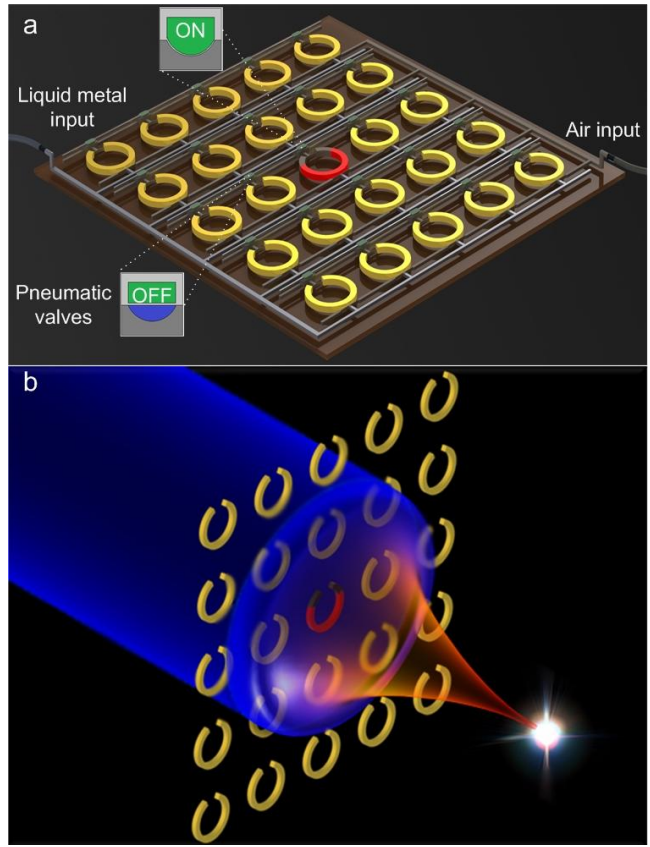

**Figure 1 | Random Access Metamaterials (RARM) and tunable flat lens. a,** The RARM is formed by loading liquid metal in to the ring micro cavities fabricated in transparent dielectric. The gaps of metallic split rings can be controlled individually in each metamolecule by regulating air pressure in the pneumatic micro cannels connected to the cavities; **b,** The randomly addressable metamaterial can be used as a flat lens with tunable focal distance when resonant properties of a split ring in the array are altered by changing the metal filling fraction.

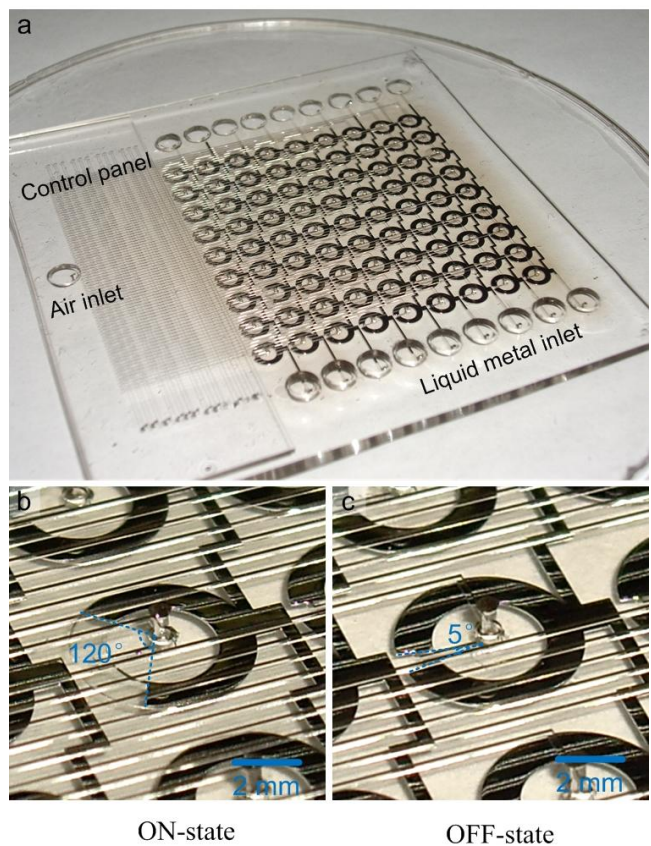

ON-state      OFF-state

**Figure 2 | Realization of the microwave RARM**. **a,** a photograph of the microfluidic chip, a square 9 × 9 array of split rings partially filled with the liquid metal, micro-channels pneumatic array and a ternary valve multiplexer addressing 81 metamolecules in the array individually at will; **b,** ON-state of one single pneumatic valve when air pushes mercury away from the cavity void forming the air gap of approximately $120^{o}$; **c,** The gap of the SRR restored to $5^{o}$ when air pressure in reduced.

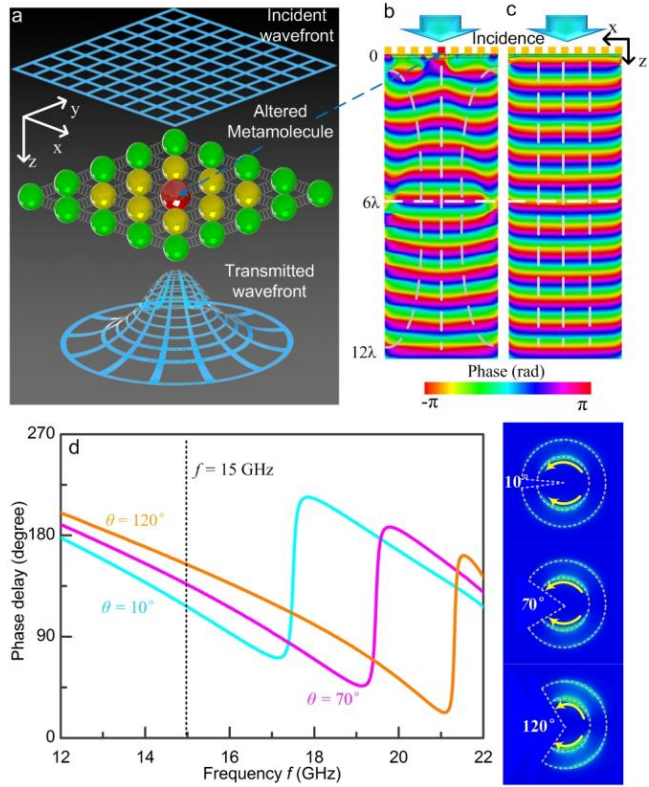

**Figure 3 | RARM and flat lens**. **a**, The schematic showing that the incident wavefront before and after transmitting trough the random access metamaterials providing functionality of a convex lens. The altered metamolecule affects the adjacent metamolecules by dipole-coupled interaction; **b**, **c** phase profile within the xz plane with and without the gap tuning of the central metamolecule. The gap $\theta$ of the central metamolecule is changed to 45 ° while the rest of the metamolecules have the gap of 5 °. **d** The phase delay inflicted by arrays of metamolecules with different gaps $\theta$. The cyan, magenta and orange lines represent the phase delay spectrum of the metamolecules array with gap widths of 10 °, 70 °, and 120 ° respectively. The inserts on the right shows the electron surface currents induced by the incident wave for gap 10 °, 70 °, and 120 ° respectively.

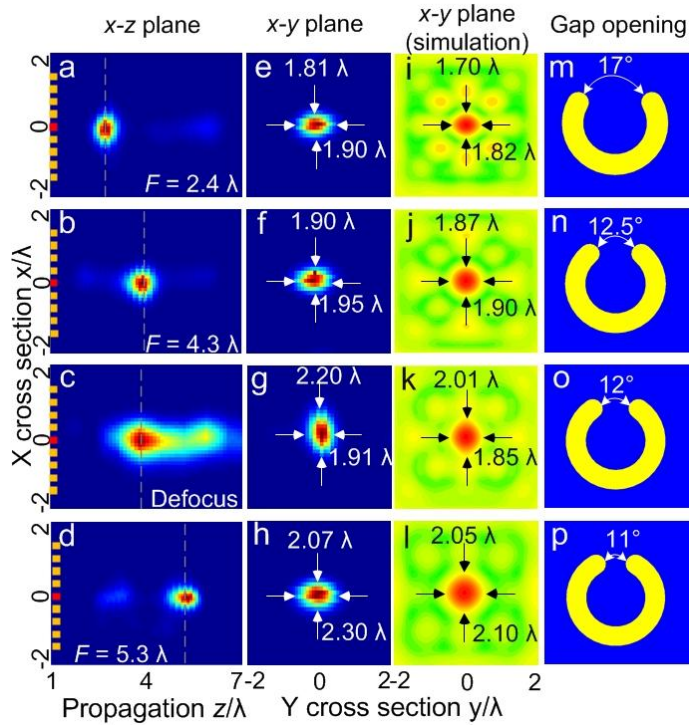

**Figure 4 | The change of the RARM lens hotspot when the gap opening is tuned. a - d** show measured longitudinal cross-sections of the hotspots created by the RARM at *xz* plane. **e – f** show the the *xy* cross section of the hotspots at the focus point while **i – l** show the corresponding simulation results. **m - p** show the gap openings of the central metamolecule are decreasing from 17 ° to 11 °. In all cases, measurements are taken with the incident wave at 15 GHz. The incident electric field is parallel to the symmetry axis of the split ring resonators within the *xy* plane and propagating along the z direction.

**Supplementary Information**

# A Random Access Reconfigurable Metamaterial and a Tunable Flat Lens

W. M. Zhu[1], Q. H. Song[2], A. Q. Liu[1†], D. P. Tsai[3], H. Cai[4], Z. X. Shen[1], R. F. Huang[5], S. K. Ting[5], Q. X. Liang[2], H. Z. Liu[2], B. H. Lu[2] and N. I. Zheludev[6,7†]

[1]*School of Electrical and Electronic Engineering*
*Nanyang Technological University, Singapore 639798*
[2]*School of Mechanical Eengineering, Xi'an Jiangtong University, Xi'an 710049, China*
[3]*Department of Physics, National Taiwan University, Taiwan 10617*
[4]*Institute of Microelectronics, A*START, Singapore 117686*
[5]*Temasek Laboratories, 5A Engineering Drive 1, Singapore 117411*
[6]*Optoelectronics Research Centre, Southampton SO17 1BJ, United Kingdom*
[7]*Centre for Disruptive Photonic Technologies, School of Physical and Mathematical Sciences*
*Nanyang Technological University, Singapore 637371*
([†] Email: eaqliu@ntu.edu.sg )

**Fabrication of the metamaterial and ternary-coded control system**

The random access reconfigurable metamaterial is a square array of $9 \times 9$ ring-shaped cavities with ring radii of 2 mm and channel width of 1 mm. The lattice constant of the array is 7 mm. It is fabricated using the polymer soft lithography technique. The micro-channels pneumatic array and a ternary valve multiplexer are fabricated by replica molding of three masters with different patterns for the metamolecules array, air pumping channels, and control panel layers followed by plasma boning of the layers with a 6-inch Pyrex glass wafer (Fig 2).

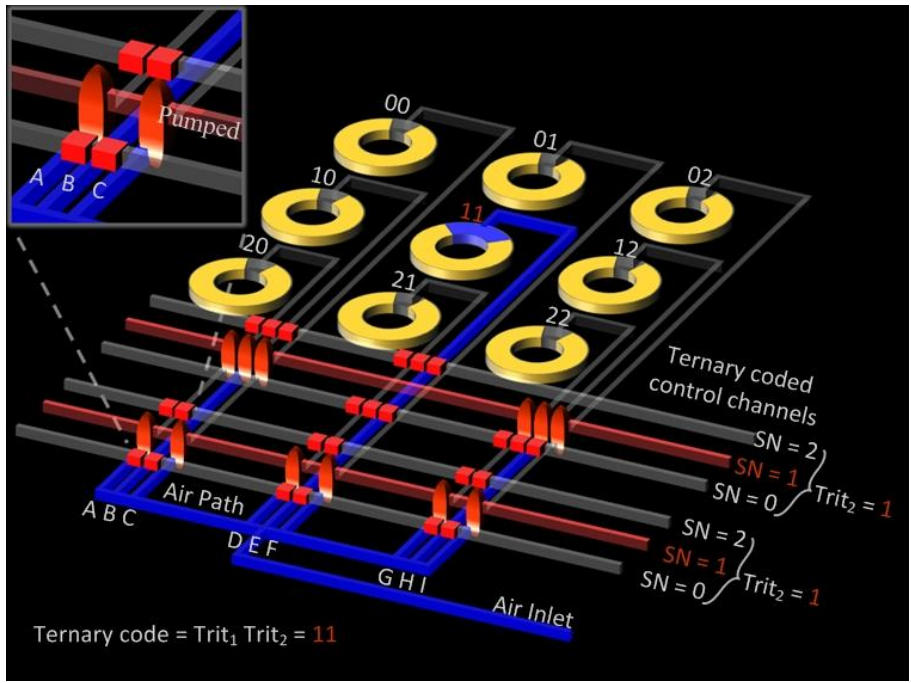

**Figure S1 | The ternary-coded pneumatic control system of the random access reconfigurable metamaterial**. The schematic shows a simplified $3 \times 3$ metamoleculars array where supply of air from the inlet to metamolecules is controlled by vales (see details in the inset) that are driven from 6 ternary coded pneumatic control channels arranged in two trits ($Trit_1$ and $Trit_2$). Only one control channel from each trit is pumped at a time giving $3^2 = 9$ combinations necessary to address each metamolecule of the entire $3 \times 3$ array at will. Metamolecules are numbered according to their x & y positions in the array. On the example given $Trit_1$ and $Trit_2$ are both equal to 1 addressing metamolecule at the x = 1, y = 1 position. In actual realization of the metamaterials to control $9 \times 9$ the entire metamolecular array, 4 trits of pneumatic control channels were used.

**Numerical analysis of the lens function of the RARM metamaterials**

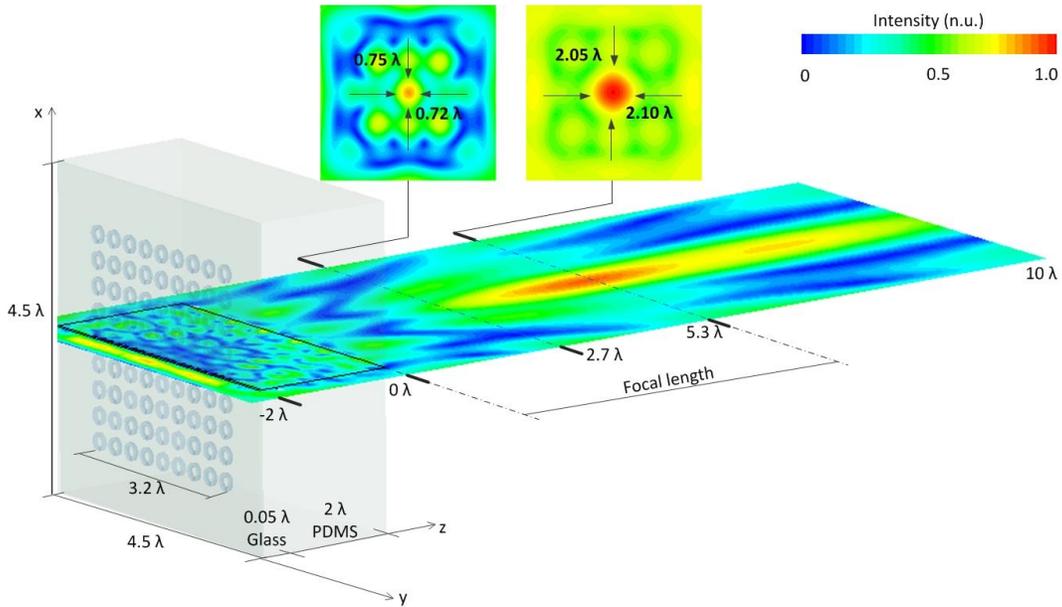

**Figure S2 | Numerical simulation of the hotspots created by the reconfigurable metamaterial.** The insets on the left and right show the *xy* cross sections at the edge and in the middle of the hotspot, respectively.

The numerical simulations are conducted by using CST microwave studio. Here, the ring gap of the central disturbed metamolecule is 10 °, while gaps in the rest of the rings are kept at 5 °. The incident plane wave propagating along the z direction has a frequency of 15 GHz and is polarized along x direction. The glass substrate (1 mm in thickness) and PDMS layer (4 cm in thickness) is considered as homogenous bulk materials with a size of 9 cm × 9 cm, which is the size of the shutter used in our experiment. To simplify the calculations, the microfluidic channels are not included in the numerical model.

The Maxwell simulations show that the hotspot is localized at 5.3 λ from the RARM and has a size of 2.05 λ × 2.10 λ. The electric field distribution is complex in the intermediate zone between the metamaterial and the hotspot. At a distance of 2.7 λ from the metamaterial, the main hotspot on the *xy* cross section has a waist size of 0.75 λ × 0.72 λ, which is smaller than the waist size of the hotspot at the focal point. However, this sub-wavelength intensity distribution is on the *xy* plane located at the edge of the focal spot, which is not the sub-wavelength focusing of the RARM lens.